\documentclass[traditabstract]{aa}
\usepackage[utf8]{inputenc}
\usepackage{amsmath}
\usepackage{amssymb}
\usepackage{natbib}
\usepackage{graphicx}
\usepackage{txfonts}
\usepackage{color}
\usepackage{multirow}
\usepackage{xspace}
\usepackage{ulem}
\bibpunct{(}{)}{;}{a}{}{,} 

\def\arcsec{\hbox{$^{\prime\prime}$}}
\def\flux{erg\,s$^{-1}$\,cm$^{-2}$}

\def\lum{erg s$^{-1}$}

\def\nustar{\textit{NuSTAR}\xspace}
\def\xmm{\textit{XMM-Newton}\xspace}
\def\swiftxrt{\textit{Swift}/XRT\xspace}
\def\nicer{\textit{NICER}\xspace}

\def \caltech {{Division of Physics, Mathematics and Astronomy,
California Institute of Technology, Pasadena, CA 91125, USA}}

\newcommand{\ero}{{eROSITA}\xspace}
\newcommand{\art}{{ART-XC}\xspace}
\newcommand{\mart}{{\it Mikhail Pavlinsky }{ART-XC}\xspace}
\newcommand{\srg}{{\it SRG}\xspace}
\newcommand{\srga}{\object{SRGA\,J204318.2+443815}\xspace}
\newcommand{\srge}{\object{SRGe\,J204319.0+443820}\xspace}
\newcommand\target{ZTF18abjpmzf\xspace}

\begin{document}

\title{\srg/ART-XC discovery of \srga: towards the complete population of faint X-ray pulsars}
\author{A.A. Lutovinov\thanks{aal@iki.rssi.ru}\inst{1} \and {S.S. Tsygankov}\inst{1,2} \and I.A. Mereminskiy\inst{1} \and S.V. Molkov\inst{1} \and A.N. Semena\inst{1} \and V.A. Arefiev\inst{1}  \and I.F. Bikmaev\inst{3} \and A.A. Djupvik\inst{4,5} \and M.R. Gilfanov\inst{1,6} \and D.I. Karasev\inst{1} \and I.Yu. Lapshov\inst{1} \and P.S. Medvedev\inst{1} \and A.E. Shtykovsky\inst{1} \and R.A. Sunyaev\inst{1,6} \and A.Yu. Tkachenko\inst{1} \and S. Anand\inst{7} \and M.C.B. Ashley\inst{8} \and K. De\inst{9} \and M.M. Kasliwal\inst{7} \and S.R. Kulkarni\inst{7} \and J. van Roestel\inst{7} \and Y. Yao\inst{7}}
\titlerunning{Discovery of new X-ray pulsar \srga}
\authorrunning{Lutovinov et al.}
\institute{
Space Research Institute (IKI) of Russian Academy of Sciences,
Prosoyuznaya ul 84/32, 117997 Moscow, Russian Federation
\and
Department of Physics and Astronomy, FI-20014 University of Turku,  Finland
\and
Kazan Federal University, Kremlevskaya Str., 18,  Kazan, Russian Federation
\and
Nordic Optical Telescope, Apartado 474, 38700 Santa Cruz de La Palma, Santa Cruz de Tenerife, Spain
\and
Department of Physics and Astronomy, Aarhus University, NyMunkegade 120, DK-8000 Aarhus C, Denmark
\and
Max Planck Institute for Astrophysics, Karl-Schwarzschild-Str. 1,
Postfach 1317, D-85741 Garching, Germany
\and
\caltech
\and
School of Physics, University of New South Wales, Sydney NSW 2052, Australia
\and
Cahill Center for Astrophysics, California Institute of Technology, Pasadena, CA 91125, USA}

\keywords{pulsars: individual: (\srga) – stars: neutron – X-rays: binaries}

\abstract{We report a discovery of a new long-period X-ray pulsar \srga/\srge in the Be binary system. The source was found in the second all-sky survey by the \mart telescope on board the \srg\ mission. The follow-up observations with \xmm, \nicer and \nustar observatories allowed us to discover a strong coherent signal in the source light curve with the period of $\sim742$ s. The pulsed fraction was found to depend on the energy increasing from $\sim20$\% in soft X-rays to $>50$\% at high energies, as it is typical for X-ray pulsars. The source demonstrate a quite hard spectrum with an exponential cutoff at high energies and bolometric luminosity of $L_X \simeq 4\times10^{35}$\,\lum. Dedicated optical and infrared observations with the {\it RTT-150}, {\it NOT}, {\it Keck} and {\it Palomar} telescopes revealed a number of emission lines (H$_{\alpha}$, \ion{He}{I}, Pashen and Braket series) with the strongly absorbed continuum. All of above suggests that \srga/\srge is a new persistent low luminosity X-ray pulsar in a distant binary system with a Be-star of the B0-B2e class. Thus the \srg\ observatory allow us to unveil the hidden population of faint persistent objects including the population of slowly rotating X-ray pulsars in Be systems.}

\maketitle

\section{Introduction}

The key task of Spectrum Roentgen Gamma ({\it SRG}) mission \citep{sunyaev21} is the deepest all-sky survey in X-rays, both at soft energies 0.3-10 keV with the \ero\ telescope \citep{2021A&A...647A...1P} and the hard ones 4-30 keV with the {\it Mikhail Pavlinsky} \art\ telescope \citep{2021A&A...650A..42P}.
Surveying about 1\% of the sky daily, down to the fluxes of $\simeq(1-2)\times10^{-11}$ erg cm$^{-2}$ s$^{-1}$ in the 4-12 keV band (about 1-1.5 mCrab), the \art\ telescope is  providing a unique possibility to study the population of faint transients that would be otherwise missed, being too weak for the wide field-of-view telescopes and all-sky monitors (like {\it INTEGRAL}/IBIS, {\it Swift}/BAT, {\it MAXI}). One of the most intriguing families of such a population are accreting neutron stars with very strong magnetic fields (X-ray pulsars) allowing studies of an interaction of the matter with the magnetic field at very low mass accretion rates.

On Nov 20, 2020 during the second consecutive all-sky survey, the {\it Mikhail Pavlinsky} ART-XC telescope discovered a relatively bright X-ray source with coordinates of RA$=310.8259$, Dec$=44.6374$ (J2000)
and the flux of ${\sim}2.4\times10^{-11}$\,\flux\ in the 4-12 keV energy band \citep{2020ATel14206....1M}.
The source was also detected by \ero\ that allowed  to obtain an enhanced position of the source
(RA$=310.8293$, Dec$=44.6390$, J2000, 95\% confidence radius is 4\arcsec) and to perform a preliminary spectral analysis in soft X-rays \citep{2020ATel14206....1M}. The \ero\ position is found to be consistent with the optical transient ZTF18abjpmzf in turn identified with a distant star at $\sim8.0^{+2.8}_{-1.9}$ kpc using the {\it Gaia} survey \citep{2021AJ....161..147B}. This distance value we will use in the following calculations.

The source ZTF18abjpmzf has been monitored by the Zwicky Transient Facility (ZTF; \citealt{Bellm2019b, Graham2019}) since March 2018. The median of all ZTF detection positions is (R.A., Dec)~$=$~($310.827745$, $44.638897$) = (20h43m18.66s, $\rm +44d38m20.0s$) (J2000), that corresponds to the Galactic coordinates ($l$, $b$)~$=$~($83.9837$, $1.3407$). The optical source is coincident with a bright infrared star detected in 2MASS \citep{Skrutskie2006}. This star exhibited a variable infrared emission in the $J$-band based on data taken with the Palomar Gattini-IR survey (PGIR; \citealt{Moore2019, De2020a}) since November 2018.

Follow-up optical observations as well as an inspection of the archival data revealed a strong H$\alpha$ line in the optical spectrum and optical/IR variability \citep{2020ATel14232....1Y,2020ATel14234....1D}.
Given these measurements as well as a low Galactic latitude of the transient and strong absorption in its spectrum we could assume that \srga\ is a new high-mass X-ray binary (HMXB) system, harboring a neutron star and a Be companion.

In this paper we report a discovery of a new transient \srga\ and results of its follow-up observations in X-rays with \xmm, \nustar, \swiftxrt and \nicer\ and optics with the RTT-150, NOT, Keck and Palomar telescopes. These observations allow to discover coherent pulsations with the period of $\sim742$\,s in
the source light curve, measure its broadband spectrum and reveal a Be-nature of the optical counterpart.

\section{Observations and data analysis}

\subsection{X-ray observations}

The Spectrum Roentgen Gamma ({\it SRG}) mission is a project of the Russian Federal Space program that hosts two X-ray telescopes: the {\it Mikhail Pavlinsky} ART-XC and \ero. The {\it Mikhail Pavlinsky} ART-XC telescope is a grazing incidence focusing X-ray telescope which provides imaging, timing and spectroscopy in the 4-30 keV energy range \citep{2021A&A...650A..42P}. The telescope consists of seven identical modules with the total effective area of $\sim450$\,cm$^2$ at 6 keV, angular resolution of $\sim50$\arcsec, energy resolution of $1.4$ keV at 6 keV and timing resolution of 23 $\mu$s. ART-XC data were processed with the analysis software {\sc artproducts} v0.9 with the {\sc caldb} version 20200401. During the second \srg\ survey \srga\ was observed with a total exposure of 125 s (an effective exposure time of $\sim37$ s). Taking into account the faintness of the source, short exposure and limited number of photons we attributed all counts detected by ART-XC into two broad energy bins (4-12 an 12-26 keV) and modeled the ART-XC data jointly with the contemporary \ero spectrum. \ero\ detected the source \srge\ in both surveys: at the moment of its discovery with \art\ in eRASS2 (Nov 18-21, 2020) with a total exposure of 252 s and (based on the retrospective analysis) in eRASS1 (May 18-22, 2020), where it was observed with a total exposure of 289 s. \ero data were processed with the \ero pipeline at IKI based on eSASS package components and data analysis software developed  at IKI. \srga/\srge was also significantly detected by \art and \ero during third survey on May 21, 2021 with the total exposures of 138 and 227 s, respectively.

\begin{figure}
    \centering
    \vbox{
    \includegraphics[width=0.95\columnwidth,bb=5 260 570 720,clip]{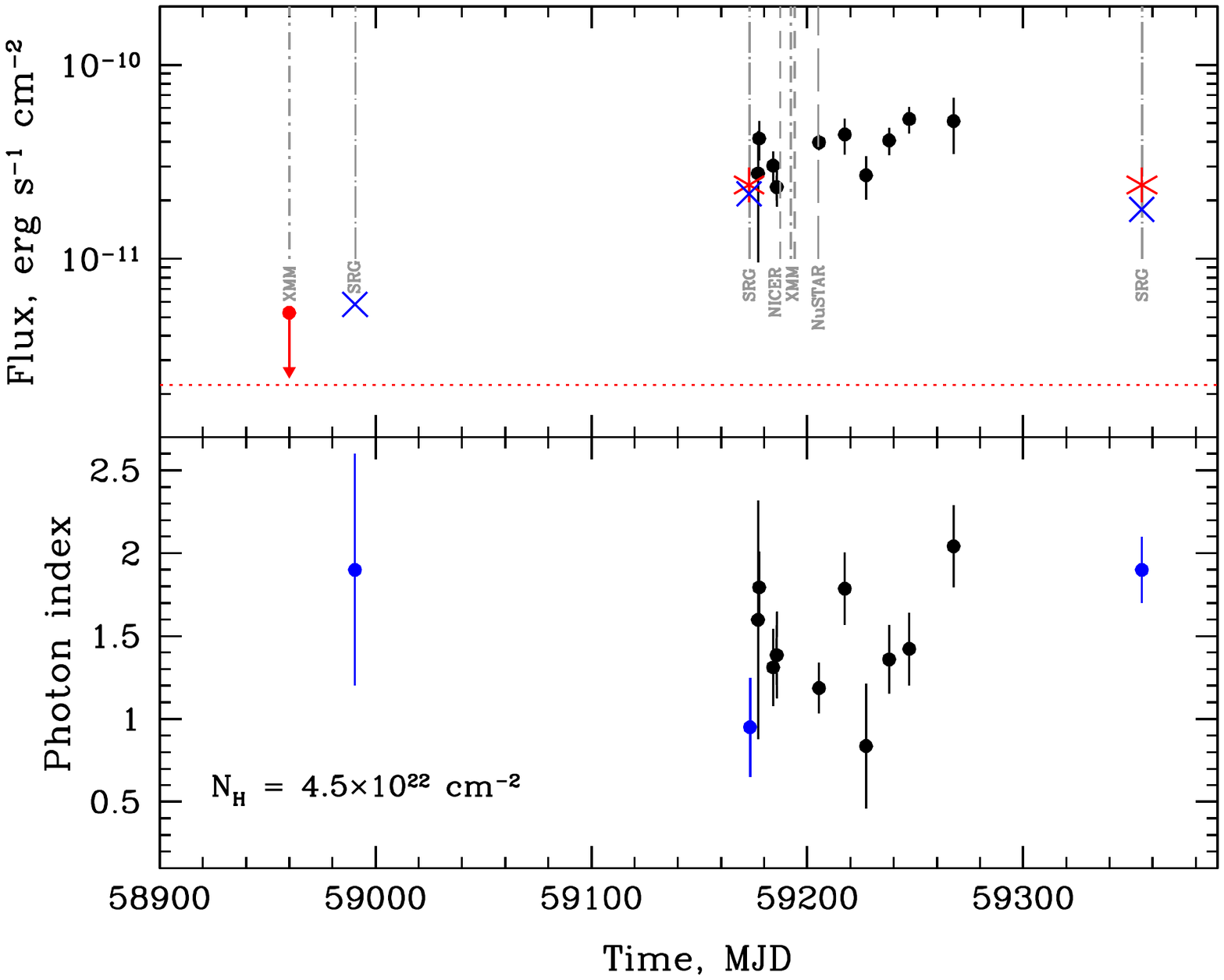}
    \includegraphics[width=\columnwidth,bb=-20 0 590 430,clip]{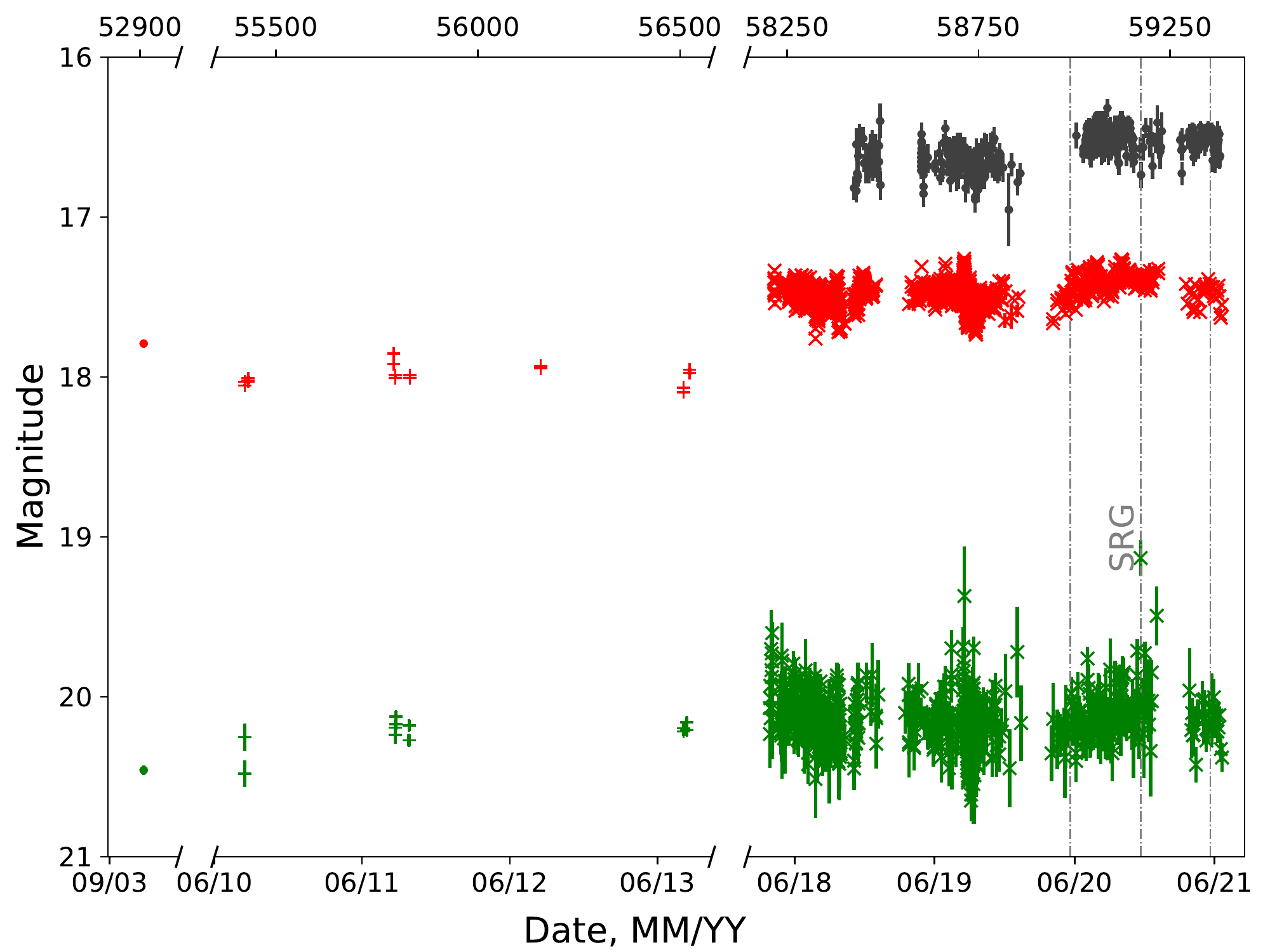}
    }
    \caption{Top panel. Light curve of \srga obtained with \swiftxrt (black points), \ero (blue crosses) in the 0.5-10 keV band and \art\ (red crosses) in the 4-12 keV band. Middle panel shows a corresponding dependence of the photon index assuming a simple absorbed power law spectral model with $N_{\rm H}$ fixed at the value of $4.5\times10^{22}$~cm$^{-2}$, determined from the average spectrum. Times of the source observations with other X-ray instruments are shown with vertical grey lines: \srg (long dashed-dotted), {\it NICER} (dashed), \xmm (dashed-dotted), \nustar (long dashed). The source flux measured by \art\ and \ero during the discovery and in the first and third surveys are shown with the red asterisk and blue crosses, respectively. Upper limits for the source flux from the \xmm\ slew surveys in Apr 2020 and Dec 2011 are shown by the red circle and red dotted line, respectively.
    Bottom panel: long-term optical light curve of \target in $g$ ($r$) filters shown in green (red). Earliest measurements (dots) are from SDSS, PanSTARRS data shown with pluses and ZTF with crosses. Also, $J$-filter measurements from $PGIR$ (shifted for clarity by $+4$ magnitudes) is shown with black dots. Secular long-term brightening is clearly seen in $r$ filter measurements, along with irregular short-scale low-amplitude variability.}
    \label{fig:xrtopt_lc}
\end{figure}

Soon after the detection of the source we initiated \nicer follow-up observations in order to examine its short-scale variability. Due to different reasons no observations were carried out until Dec 04, 2020, when first short (900 s) observation was performed. Later, more longer observation was carried out on Dec 11, 2020, which lasted for 8 ks. We used a standard {\sc NICERDAS} in order to extract spectra and light curves. Unfortunately, the source was too obscured and weak, with energies below $\approx$2 keV and above $\approx$5 keV being dominated by a non-photon background. Therefore we decided to not use spectral data. In the 2-6 keV energy band we found a significant modulation with the period of 741.8 s, which shape is roughly consistent with measurements by \xmm.

To improve further the source localization and to study in depth its spectral an timing properties in soft X-rays, we asked follow-up DDT observations of \srga\ with the \xmm observatory. Observations were performed on Dec 9, 2020 (ObsID 0872391201). To process the {\it XMM-Newton} data, we used version 18.0 of the XMM-Newton Science Analysis System (SAS). For the analysis of the EPIC data we selected events with patterns in the range 0--4 for the pn camera and 0--12 for the two MOS cameras, using a circular region with a radius of 20\arcsec\ around the source position. Background events were selected from circular regions (with radius of 30\arcsec) offset from the source position.

To trace the long-term variability of the source observations with the X-ray telescope \citep[XRT; ][]{2005SSRv..120..165B} on-board the {\it Neil Gehrels Swift Observatory} \citep{2004ApJ...611.1005G} were triggered. This monitoring lasted for three months from November 24, 2020 to February 22, 2021.
All XRT data were taken in the photon counting mode and reduced using the online tools\footnote{\url{http://www.swift.ac.uk/user_objects/}} \citep[][]{2009MNRAS.397.1177E} provided by the UK Swift Science Data Centre. The resulting spectra were fitted with a simple absorbed power law model in the 0.3--10~keV band using the {\sc xspec} package \citep{Arn96}.

\begin{figure}
    \centering
    \includegraphics[width=0.95\columnwidth,bb=10 130 540 570,clip]{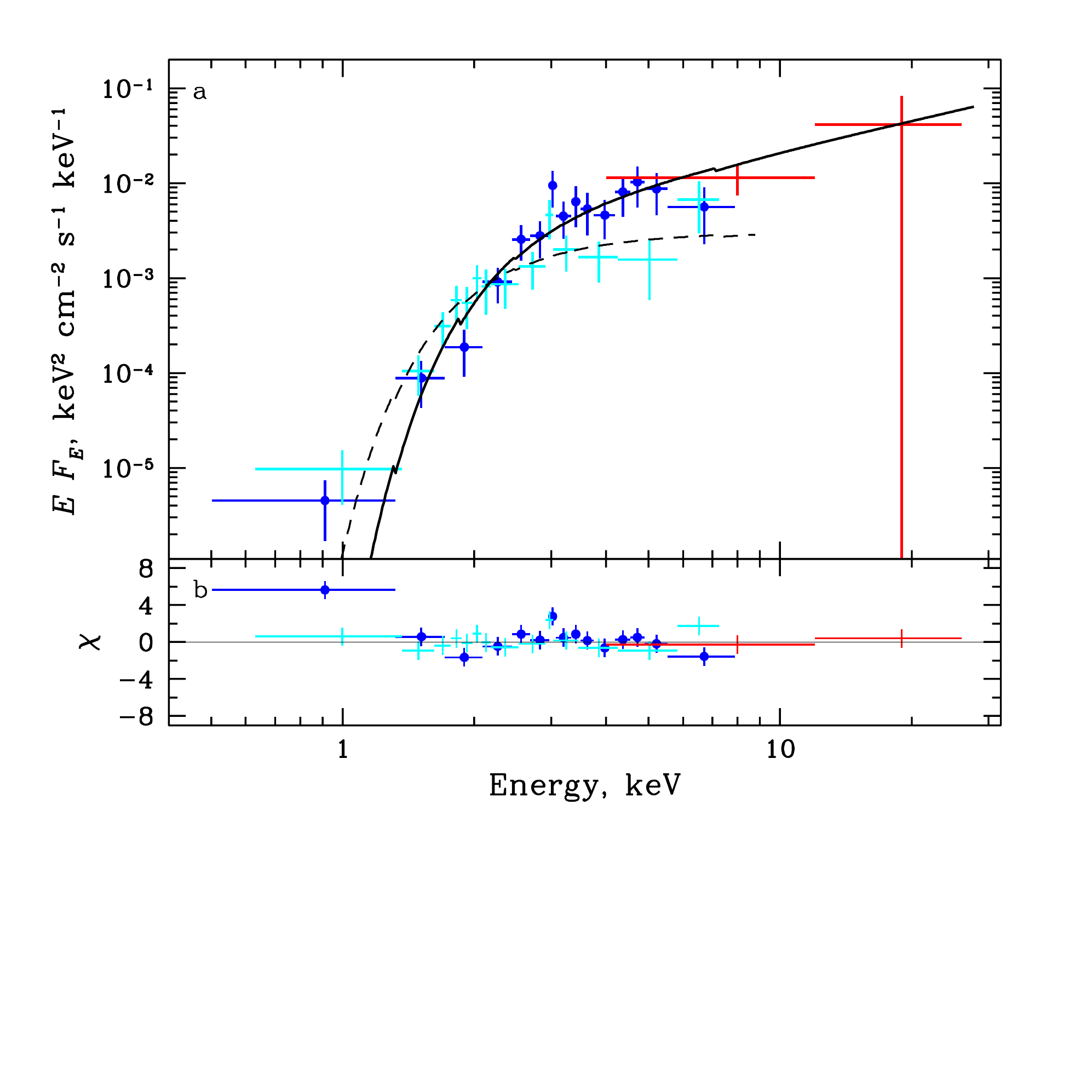}
    \caption{Spectra of \srga measured with \srg instruments. Blue and red points correspond to \ero and \art observations in the second survey, respectively. Cyan points represent the spectrum obtained by \ero during the first survey. Solid and dashed lines are corresponding best-fit models.}
    \label{fig:srgsp}
\end{figure}

Finally in our analysis we used the \nustar data to confirm detection of the pulsations, which were initially found in the \xmm and \nicer data, as well as to study properties of the source in the broad X-ray band. The {\it NuSTAR} observatory consists of two identical X-ray telescope modules, referred to as FPMA and FPMB \citep{2013ApJ...770..103H}. It provides X-ray imaging, spectroscopy and timing capabilities in the energy range of 3-79~keV. {\it NuSTAR} performed an observation of \srga on Dec 22, 2020 (ObsID 90601338002).  The data reduction for this observation was done using HEASOFT v6.28 and CalDB version 20210210.
All spectra were binned to have at least one count per energy bin and W-statistic\footnote{\url{https://heasarc.gsfc.nasa.gov/xanadu/xspec/manual/XSappendixStatistics.html}} was applied
\citep{1979ApJ...230..274W}.

\subsection{Optical and infrared data}

{\it RTT-150}. Optical observations from the Russian-Turkish Telescope (RTT-150) have been performed by using the TFOSC instrument on 24-25 Nov, 2020 and 9 Apr, 2021 (low resolution spectrometer and imager) equipped with the Andor CCD (model DZ936, BEX2 DD chip) cooled to -80 C. We used g,r,i,z filters for photometrical observations. Note, that during RTT-150 observations the optical counterpart was by 0.3-0.5 mag brighter in comparison with SDSS DR16 archive data, showing long term variations its brightness between November 2020 and April 2021. At the same time we found no fast variations in brightness at the time scale of 3 hours during RTT-150 photometrical observations on Nov 24, 2020, when we have obtained two low resolution (15 Angstrom) spectra with the exposure of 900 sec each.

\begin{figure}
    \centering
    \includegraphics[width=0.95\columnwidth,bb=10 130 540 560]{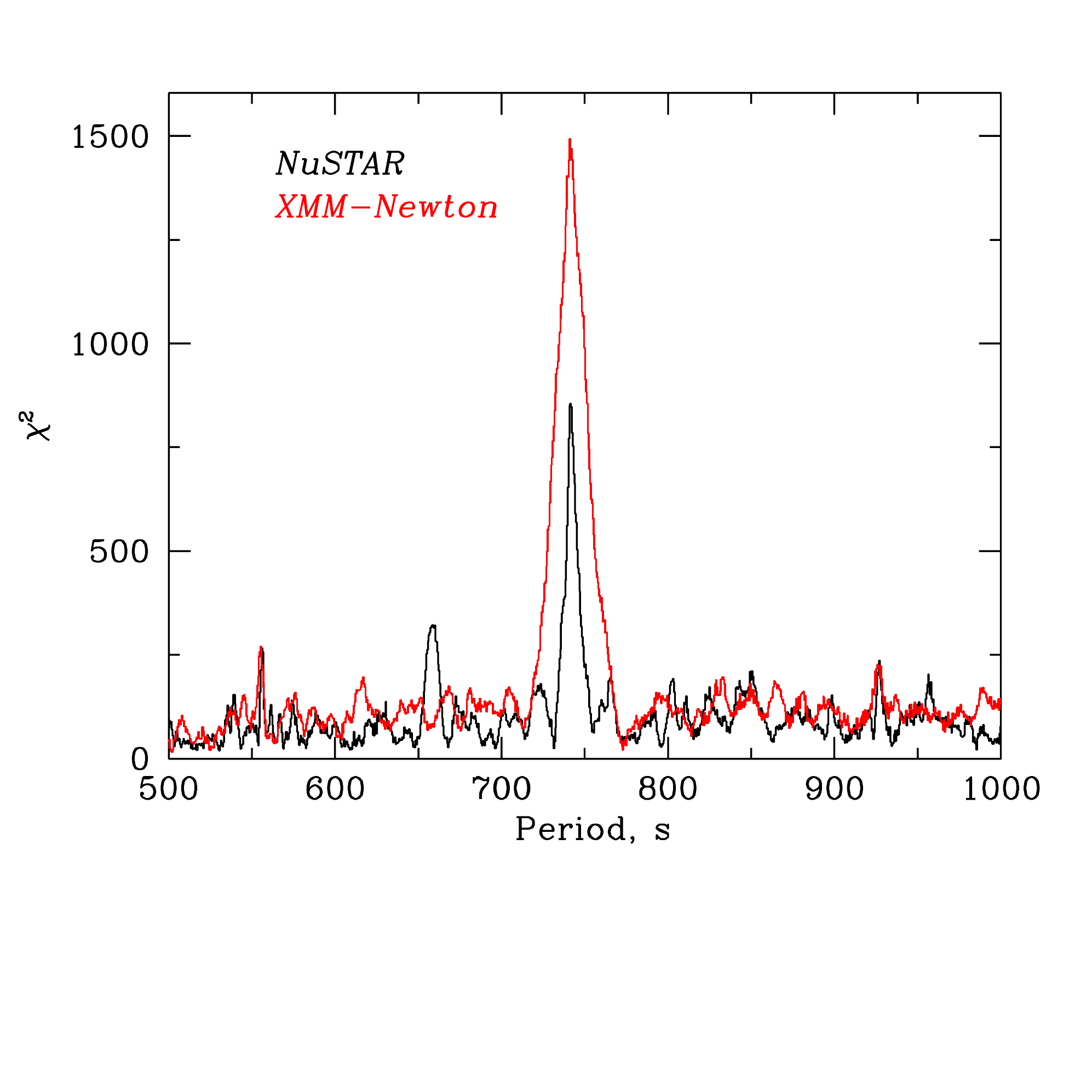}
    \caption{Periodogram of the source, obtained by \nustar (black) and \xmm (red). A coherent signal is clearly detected by both instruments around $\simeq742$ s.}
    \label{fig:efs}
\end{figure}

{\it NOT}.
At the 2.56m Nordic Optical Telescope we applied for time through the fast-track access and obtained an optical low-resolution spectrum on 2021 March 18 with ALFOSC and near-IR low-resolution spectra on April 25 with NOTCam. The ALFOSC spectrum with Grism \#4 and a 1$\arcsec$ wide slit covers
the wavelength range 3200 to 9600 \AA \ with a resolving power of R = 360.
The exposure time was 2700s. The near-IR spectra were obtained
with NOTCam Grism \#1 and the broad-band filters $J$ ad $K$ as order sorters.
The dispersions are 2.5 and 4.1 \AA/pix in J and K, respectively, and with
the 0.6$\arcsec$ wide slit give a resolving power of R = 2100. Six spectra were taken dithering the target along the slit with individual exposure times of 300 s using ramp-sampling readout modes. Each sky-subtracted and flat-fielded spectrum was optimally extracted and wavelength calibrated before combined to a final spectrum. A nearby bright star of spectral type A0 V was observed immediately before the target and used for telluric correction, removing primarily the stellar absorption lines, upon which a properly flux-scaled Vega model was multiplied back to restore the slope
and obtain a rough flux calibration. For the $J$ band we used the telluric
standard to find the conversion from ADU/s to f$_{\lambda}$, while for the
$K$ band we used the target acquisition image with comparison stars.

{\it Keck, Palomar.} We obtained follow-up optical and near-infrared spectra using the Low Resolution ($R\approx 1000$) Imaging Spectrograph (LRIS/Keck-I; \citealt{Oke1995}) on 2020 December 12, the medium-resolution ($R\approx13000$) Echellette Spectrograph and Imager (ESI/Keck-II; \citealt{Sheinis2002}) on 2020 December 08, and the medium-resolution ($R=2700$) Triplespec spectrograph (Tspec/Palomar 200-inch telescope; \citealt{Herter2008}) on 2020 December 23.

\begin{figure*}[]
    \centering
    \hbox{
    \includegraphics[width=0.95\columnwidth,bb=20 80 530 600]{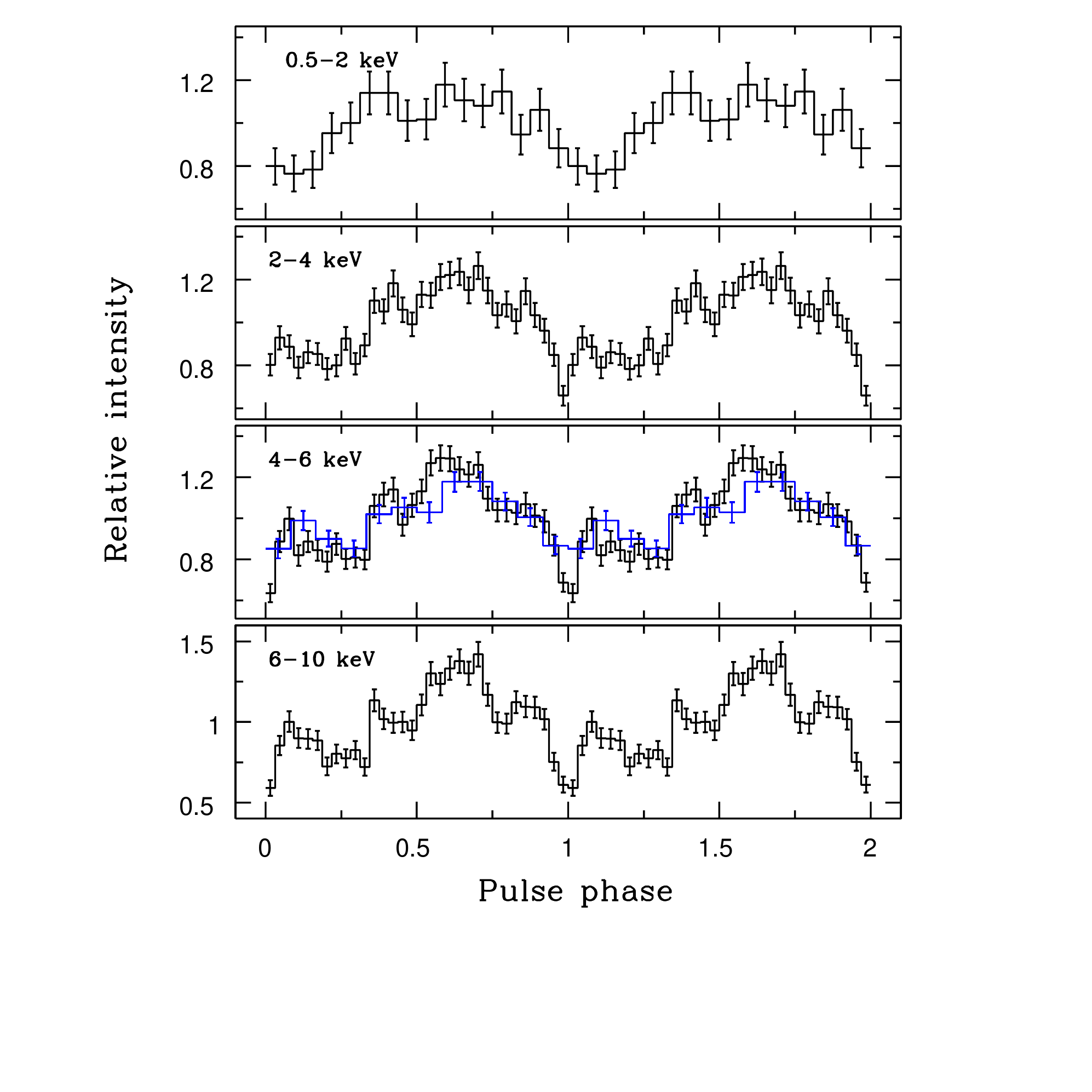}
    \includegraphics[width=0.95\columnwidth,bb=20 80 530 600]{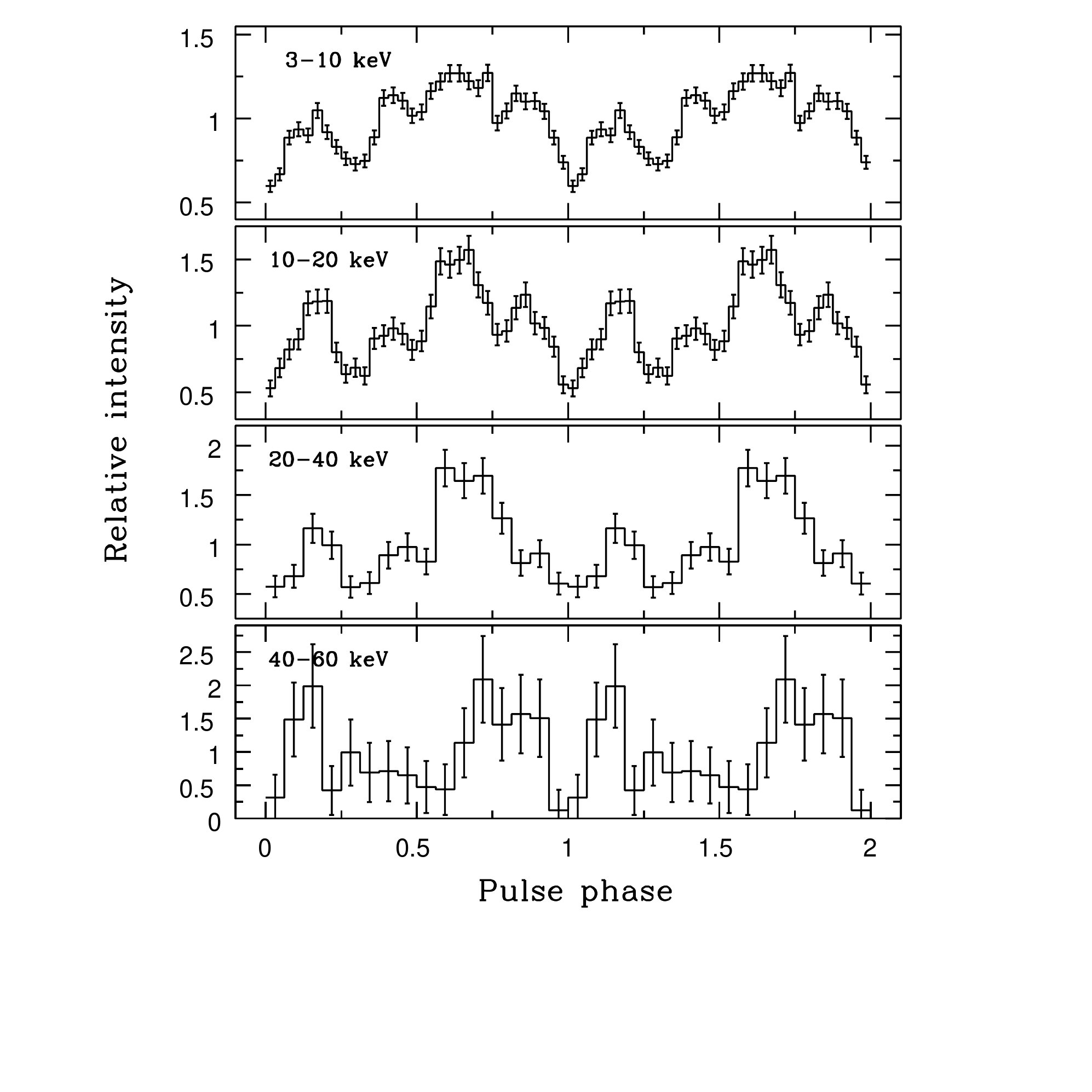}
    }
    \caption{Pulse profiles of \srga as a function of the energy as observed by \xmm (left) and \nustar (right). Profile shown in blue colour represents results obtained from the {\it NICER} data in the 2-6 keV band. Zero phase in both datasets is chosen arbitrary to coincide with the main minimum in the profile.}
    \label{fig:profile}
\end{figure*}

\section{Source discovery and long term variability}

The long-term X-ray light curve covering about 3 months of observations of \srga\ with the {\it Swift}/XRT telescope after its discovery is presented in Fig.~\ref{fig:xrtopt_lc}.  As can be seen from the figure, no significant source variability was revealed not only on this time scale, but also during a half of year after the discovery, when both \srg instruments detected again \srga in May, 2021 approximately at the same level of intensity. At the same time the X-ray source exhibited a moderate variability between the first and the second \srg all sky surveys separated by six months.

Additionally, an inspection of previous scans of this sky field during the \xmm slew surveys\footnote{\url{https://www.cosmos.esa.int/web/xmm-newton/xsa}} \citep{saxton08} (in Jun, 2010, Dec, 2011, Dec, 2018 and Apr 2020) showed no detection of the source with the lowest flux limit of $2.2\times10^{-11}$\,\flux\ in the 0.2-12 keV energy band in 2011 (red dotted line in Fig.\,\ref{fig:xrtopt_lc}). This limit points to the source flux variability at least at
the level of the order of magnitude.

As it was mentioned above the source \srga\ was initially discovered in the near-real time analysis of the \art and \ero data during the second all-sky survey. The retrospective detailed analysis of the data from both instruments obtained during the first survey showed that the source was registered by \ero while staying below the detection threshold in the \art data.
The spectra of \srga measured by \ero+ \art and \ero in the first and second surveys, respectively, are shown in Fig.\ref{fig:srgsp}. It is seen that both of them are quite similar and can be describe by an absorbed power law. The absorption values $N_{\rm H}=(3.0\pm0.8)\times10^{22}$ and $N_{\rm H}=(4.5\pm0.8)\times10^{22}$ cm$^{-2}$, measured in first and second surveys, respectively, are compatible with each other and exceed the Galactic value in the source direction. It can indicate presence of the intrinsic absorption in the system similar to already found in a number of faint BeXRB systems. The source flux in the 0.5-10 keV band obtained during the second survey $(2.2\pm0.4)\times10^{-11}$~\flux\ increased by factor of $\simeq3.7$ in comparison to the one in the first survey $5.8^{+2.3}_{-1.6}\times10^{-12}$~\flux. It was accompanied by a hardening of the spectral index from $1.9\pm0.7$ to $0.95\pm0.30$, that allowed us to register the source with the \art telescope (Fig.\ref{fig:srgsp}).

The optical counterpart of \srga \target is a nearly persistent optical source that has been monitored by optical surveys, including the Sloan Digital Sky Survey (SDSS, \citealt{Alam2015}) DR12, the Panoramic Survey Telescope and Rapid Response System (Pan-STARRS, PS1) DR1 \citep{Flewelling2020, Waters2020}, and the ZTF DR5.

We also obtained forced photometry on ZTF difference images using the publicly available tool\footnote{See details in \url{http://web.ipac.caltech.edu/staff/fmasci/ztf/forcedphot.pdf}} \citep{Masci2019}. For each band, we computed the sum of flux density in the reference image and difference images. The full optical light curve of \target is shown at the bottom panel of Fig.~\ref{fig:xrtopt_lc}.

The optical photometry from the ZTF survey allows one to search for the possible orbital period in the system. From a measured equivalent width of H$_{\alpha}$ line $EW(H_{\alpha}) = -17.40 \pm 0.07~\AA$ (see Sec.~\ref{subsec:spec}) using a relation between $EW$ and the orbital period for BeXRBs  \citep[see, e.g.][]{Reig11} we can expect an orbital period to be about $50$ days. Otherwise, from the Corbet diagram and known spin period one could estimate the \srga\, orbital period to be $\approx200$ days.

Similarly, we performed forced photometry of the PGIR20fah -- NIR counterpart of ZTF18abjpmzf -- using the PGIR images
based on the method described in \citet{De2020b}. The $J$-band light curve is shown in Fig.~\ref{fig:xrtopt_lc}.

To search for the source orbital period we calculated a generalized Lomb-Scargle periodogram \citep{scargle82} from all available ZTF {\it r}-filter measurements. We also produced a hundred simulated data sets with similar level of red noise at higher frequencies. No significant peaks are observed in periodogram.
Similar analysis was carried out for PGIR data with no prominent peaks found.

\section{Timing analysis and discovery of pulsations}

Soon after the discovery of the source with \art we triggered a TOO observation with the \xmm observatory in order to study its properties in details. The observation has been performed around 3 weeks after the discovery. As can be seen from Fig.~\ref{fig:xrtopt_lc} the source was found in the same intensity state. For the timing analysis we applied the barycentric  correction. Using standard epoch folding technique we were able to discover very strong pulsations of flux with period $P_{\rm s}=741.4\pm0.3$~s (see Fig.~\ref{fig:efs}). About two weeks after the \xmm observation \srga was observed with \nustar. Using these data we were able to confirm presence of pulsations in the source light curve with period $P_{\rm s}=741.8\pm0.1$~s. As can be seen from period values obtained from the \xmm and \nustar data, the source does not experience strong spin-up or spin-down that is compatible with the stability of its flux.
Uncertainties for the pulse period values were determined from the simulated light curves following the procedure described by \citet{2013AstL...39..375B}.

\begin{figure}
    \centering
    \includegraphics[width=0.95\columnwidth,bb=10 125 540 535,clip]{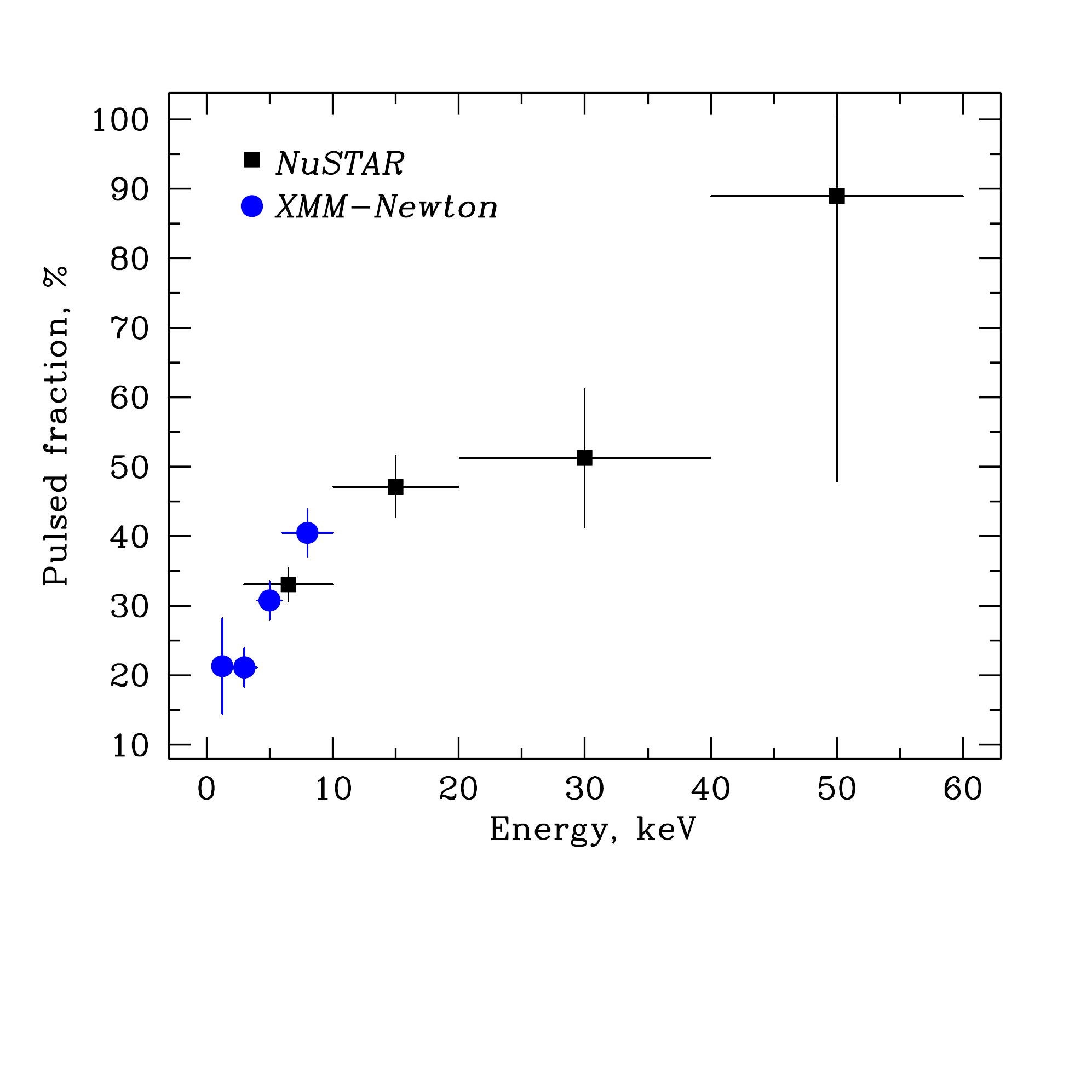}
    \caption{Pulsed fraction of \srga as a function of the energy as observed by \xmm (blue circles) and \nustar (black squares).}
    \label{fig:pulfr}
\end{figure}

Using the corresponding spin periods we constructed the pulse profile of \srga in different energy bands (Fig.~\ref{fig:profile}). The pulse profile shape demonstrates a clear dependence on the energy. At low energies (below several keV) it has simple single-peak structure. At higher energies the multiple peaks become more and more prominent. Already in the 10-20 keV band the profile consists of four evenly distributed over the spin phase peaks with a comparable intensity. At even higher energies the pulse profile is dominated by two main peaks at phases around 0.1 and 0.8.

A broad-band coverage provided by joint use of the \xmm and \nustar data allowed us also to study dependence of the pulsed fraction\footnote{defined as $\mathrm{PF}=(F_\mathrm{max}-F_\mathrm{min})/(F_\mathrm{max}+F_\mathrm{min})$, where $F_\mathrm{max}$ and $F_\mathrm{min}$ are maximum and minimum fluxes in the pulse profile, respectively} on the energy. It was found that it is increasing with the energy (Fig.~\ref{fig:pulfr}), which is typical behaviour for the majority of XRPs \citep[e.g.,][]{2009AstL...35..433L}. Moreover, the increase rate is changing around 10-15 keV, that may point to the presence of different physical mechanisms responsible for the observed emission in different energy bands, or presence of some scattering medium affecting low-energy photons.

\section{Broad-band X-ray spectrum}

\begin{figure}
    \centering
    \includegraphics[width=\columnwidth,bb=10 140 540 550,clip]{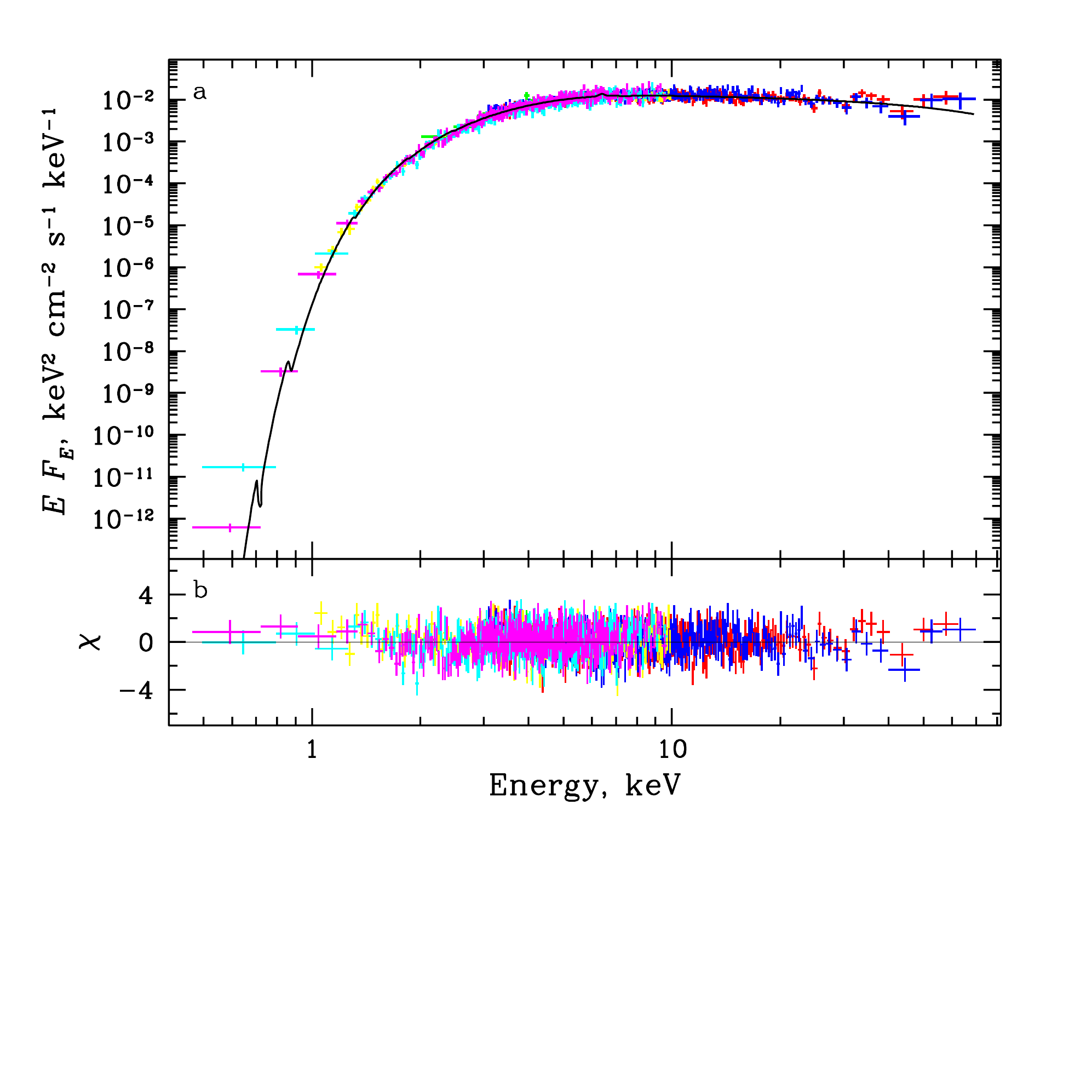}
    \caption{(a) Energy spectrum of \srga as measured by \swiftxrt (green points), \xmm (yellow, cyan and magenta points for PN, MOS1 and MOS2 instruments, respectively) and \nustar (blue and red points for FPMA and FPMB modules, respectively). Black line corresponds to the best fit model {\sc compTT+gauss+phabs}. (b) Residuals from the model.  }
    \label{fig:spe}
\end{figure}

As it follows from the \swiftxrt\ monitoring and \srg\ observations, the source flux and spectral parameters evolve only slightly. Therefore we can approximate  spectra obtained by \xmm, \nustar and \swiftxrt\ simultaneously.

As it is well known, broad band spectra of X-ray pulsars differ significantly from the spectra of neutron stars with low magnetic field in low mass X-ray binaries and from the spectra of accreting black holes \citep{1991SvAL...17..409S}.  The broadband spectrum of \srga\ is typical for accreting magnetized neuron stars and demonstrates an exponential cutoff at high energies (Fig.\,\ref{fig:spe}). Therefore, we fitted it with several continuum models usually used for such objects: a power law with a high energy exponential cutoff ({\sc highcut} or {\sc cutoffpl} in the {\sc xspec} package), combination of negative and positive power laws with exponential cutoff and comptonization model {\sc compTT}. Additionally components accounting for the interstellar absorption and fluorescent iron line at 6.4 keV were added as the {\sc phabs} model and Gaussian.

The source and background spectra from all three cameras of \xmm (MOS1, MOS2, PN), both the FPMA and FPMB modules of \nustar as well as from the \swiftxrt\ telescope were used for simultaneous fitting. To take into account the uncertainty in the instrument calibrations as well as the lack of full simultaneity of observations, cross-calibration constants between them were included in the spectral model.

All the above mentioned models describe the source spectrum quite adequately, but the {\sc compTT} model gives the better fit quality ($\chi^2=1385$ (1279 dof) for {\sc cutoffpl}, $1353$ (1278 dof) for {\sc powerlaw*highcut}, $1326$ (1278 dof) for {\sc compTT}). The best-fit parameters are summarized in Table\,\ref{tab:spe}.

There are no obvious absorption features connected with cyclotron resonant scattering features in the range 5-50 keV, that allowed us to roughly constrain a magnetic field strength of the neutron star as $B<5\times10^{11}$ or $B>5\times10^{12}$\,G.

The conclusion about an absence of additional spectral features was also verified using the phase-resolved spectroscopy, as there are pulsars in which the cyclotron line or its higher harmonics are appeared only at certain phases of rotation of the neutron star \citep[see, e.g.][]{2019ApJ...883L..11M,2021arXiv210609514M}. Particularly, we produced energy spectra of the source in 10 evenly distributed phase bins. Unfortunately, low counting statistics didn't allow us to detect significant variations over the pulse of any spectral parameters of our best fit model.

\begin{table}[]
\small
        \begin{center}
        \caption{Best-fitting results for the \srga spectrum}
        \label{tab:spe}
        \begin{tabular}{lc}
\hline
\hline
  Parameter$^{a}$  & Value \\
\hline
$T_0$, keV             &  $1.51\pm0.03$ \\
$T_{\rm p}$, keV       &  $24.6^{+44.0}_{-7.7}$ \\
$\tau_{\rm p}$         &  $1.16^{+0.55}_{-0.81}$ \\
$N_{\rm H}$            &  $4.48\pm0.11$  \\
$E_{\rm Fe}$, keV      &  $6.38\pm0.05$  \\
$\sigma_{\rm Fe}$, keV &  $0.12\pm0.06$  \\
$C_{\rm B}$            &  $1.071\pm0.015$ \\
$C_{\rm XRT}$          &  $1.083\pm0.064$ \\
$C_{\rm MOS1}$         &  $1.008\pm0.015$ \\
$C_{\rm MOS2}$         &  $0.911\pm0.014$ \\
$C_{\rm PN}$           &  $0.914\pm0.012$ \\
$F_{\rm X}$, \flux  &  $5.08\times10^{-11}$  \\
$\chi^2$ (d.o.f.)    &  1325.8 (1278) \\
\hline
    \end{tabular}
    \end{center}
        $^{a}$ Here $T_{\rm p}$, $\tau_{\rm p}$ and $T_0$ are the plasma temperature, plasma optical depth, temperature of the seed photons for the {\sc comptt} model. Fluxes are given  in the 0.5--100 keV energy range.
\end{table}

\section{Optical observations}

\subsection{Optical and Infrared (OIR) Photometry}

\begin{figure}
    \centering
    \includegraphics[width = 0.95\columnwidth]{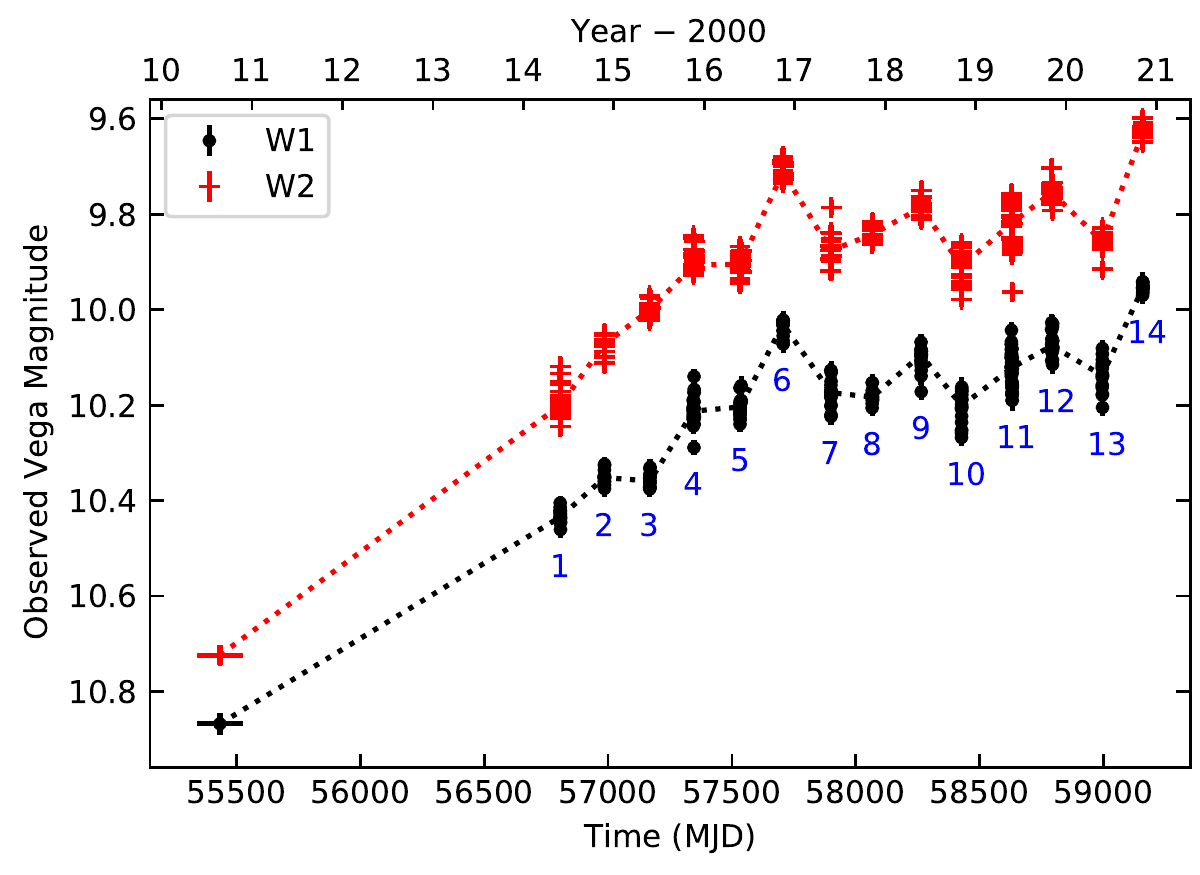}
    \includegraphics[width = 0.95\columnwidth]{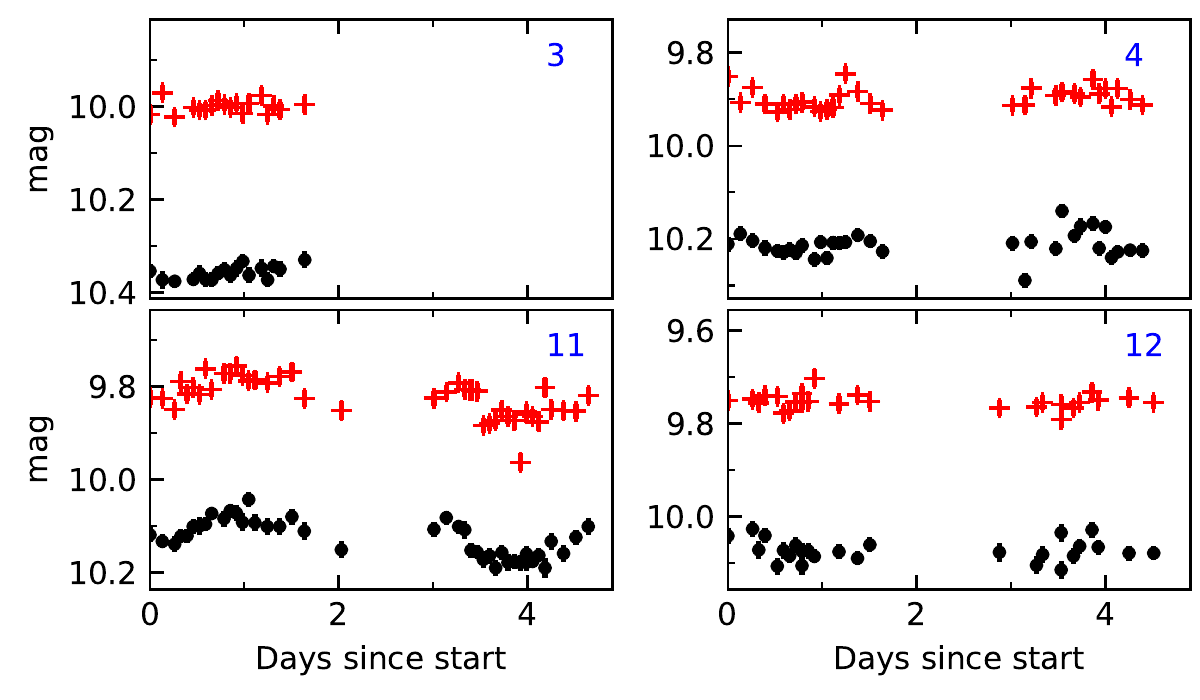}
        \caption{WISE light curve of \target in $W1$ and $W2$ bands.
        The lower panels show the zoom in of four NEOWISE visits.
        \label{fig:NEOWISElc}}
\end{figure}

\target has been observed by the Two Micron All-Sky Survey (2MASS; \citealt{Skrutskie2006}), the Wide-field Infrared Survey Explorer (WISE) telescope \citep{Wright2010}, and the Palomar Gattini-IR \cite{De2020Gattini}. In the AllWISE catalog (circa 2010, \citealt{Cutri2013}), the source was detected at $W1=10.868\pm 0.023$, $W2=10.724\pm 0.021$, $W3=10.326\pm 0.107$, and $W4 = 8.905\pm 0.410$ (Vega system). During the post-cryogenic phase from 2014 to 2020, it was also monitored by Near-Earth Object WISE Reactivation Mission (NEOWISE; \citealt{Mainzer2014}) every 6 months in the $W1$ and $W2$ bands. The infrared light curves are presented in Figure~\ref{fig:NEOWISElc}.

As reported by \citet{2020ATel14232....1Y} and \citet{2020ATel14234....1D}, \target exhibits a secular brightening and undulations on timescales of years in the optical and infrared bands. From 2010 to 2020, the source has clearly brightened by $\Delta g \approx 0.10$, $\Delta r \approx 0.65$, $\Delta W1 \approx 0.82$, and $\Delta W2 \approx 0.98$. \target also exhibits transient fluctuations at the level of 0.1--0.2\,mag on the timescale of a few days. These results are also confirmed by {\it RTT-150}, revealing long term variations of the object brightness at the time scale of several months (see above).
The PGIR $J$-band lightcurve exhibit long term low amplitude variability from the source in addition to a secular brightening of $\approx 0.2$\,mag between 2019 November and 2020 May.

\begin{figure*}[]
    \centering
    \includegraphics[width=0.95\textwidth]{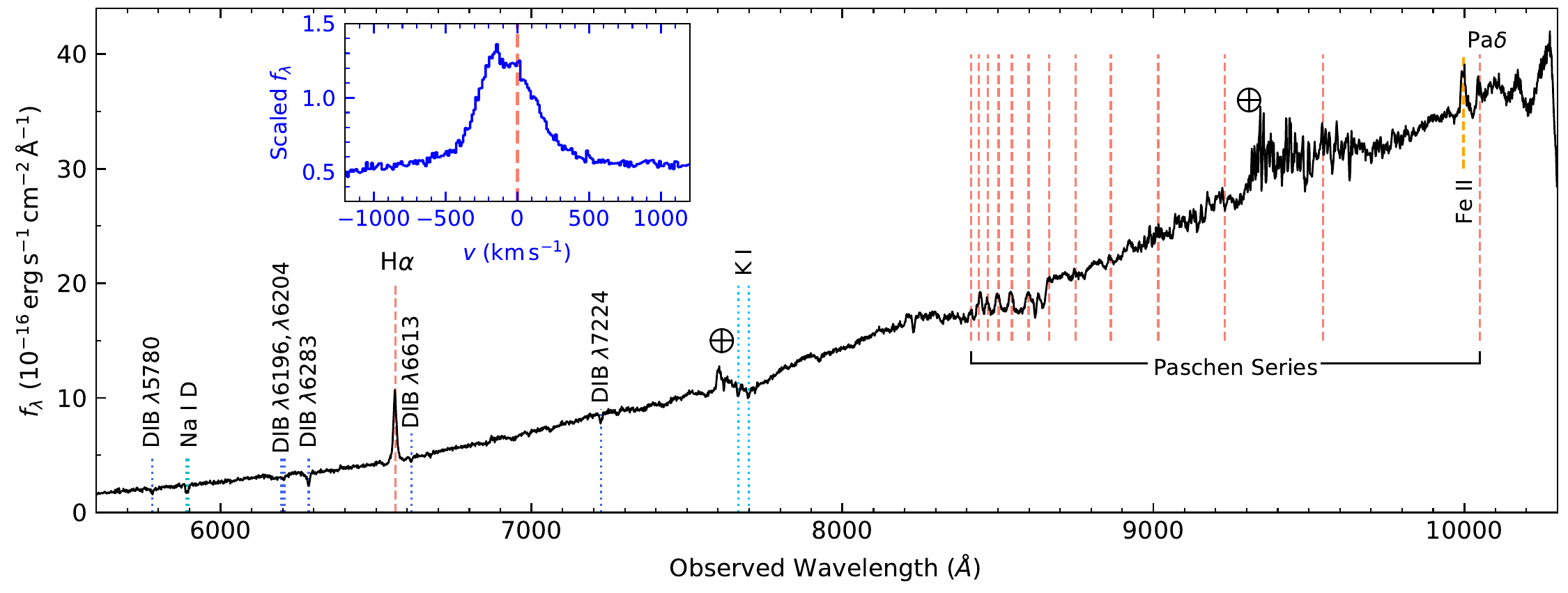}
    \includegraphics[width=0.95\textwidth]{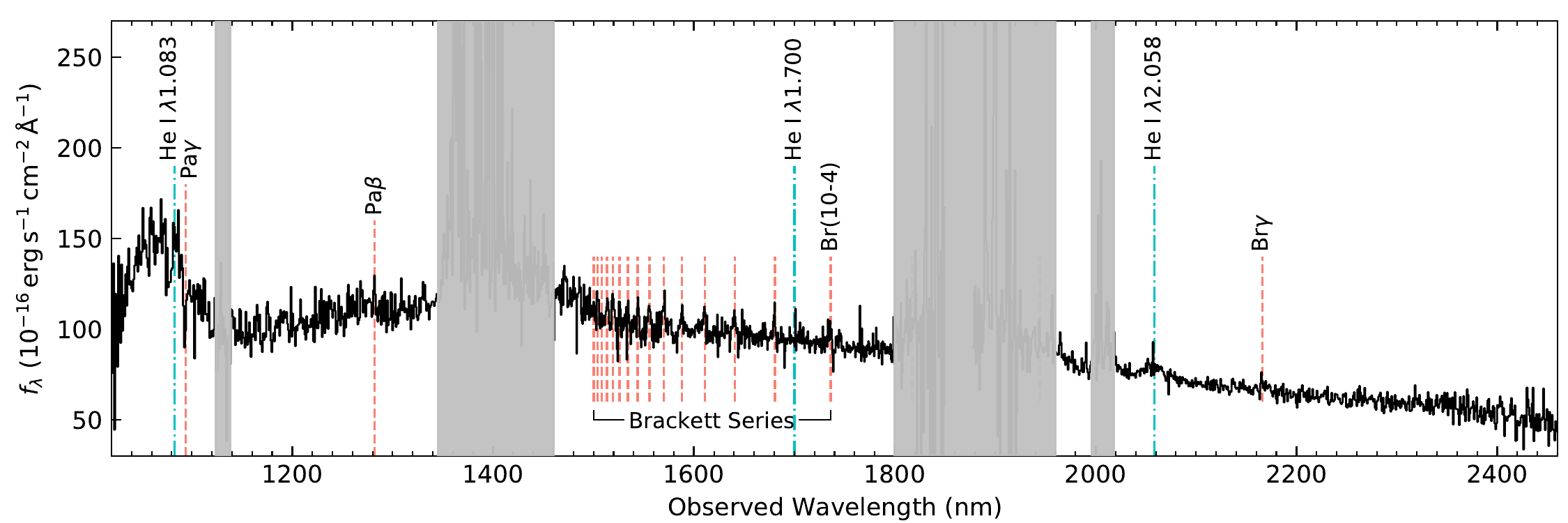}
    \caption{\textit{Upper}: Observed LRIS spectrum of \target. The inset shows the H$\alpha$ profile in the ESI spectrum, where the instrumental broadening is $\sigma_{\rm inst}=16\,{\rm km\,s^{-1}}$. \textit{Bottom}: Observed Tspec spectrum of \target. Wavelength ranges of high atmospheric opacity are masked in grey. \label{fig:spec}}
\end{figure*}

Long-term secular variations in brightness and optical-NIR colors in BeXRBs are often associated with the disk-loss episodes \citep[see, e.g.][ and references therein]{wisniewski10}. During such episodes, the optical source becomes fainter and bluer \citep{reig15photom}, while during the disk recovery the source is brighten and redden. This is closely reminiscent the observed brightness evolution of \srga\, and also agrees well with non-detection of the source in 2011 by \xmm .

\begin{figure}
    \centering
    \includegraphics[width=0.99\columnwidth,bb=10 40 740 600,clip]
    {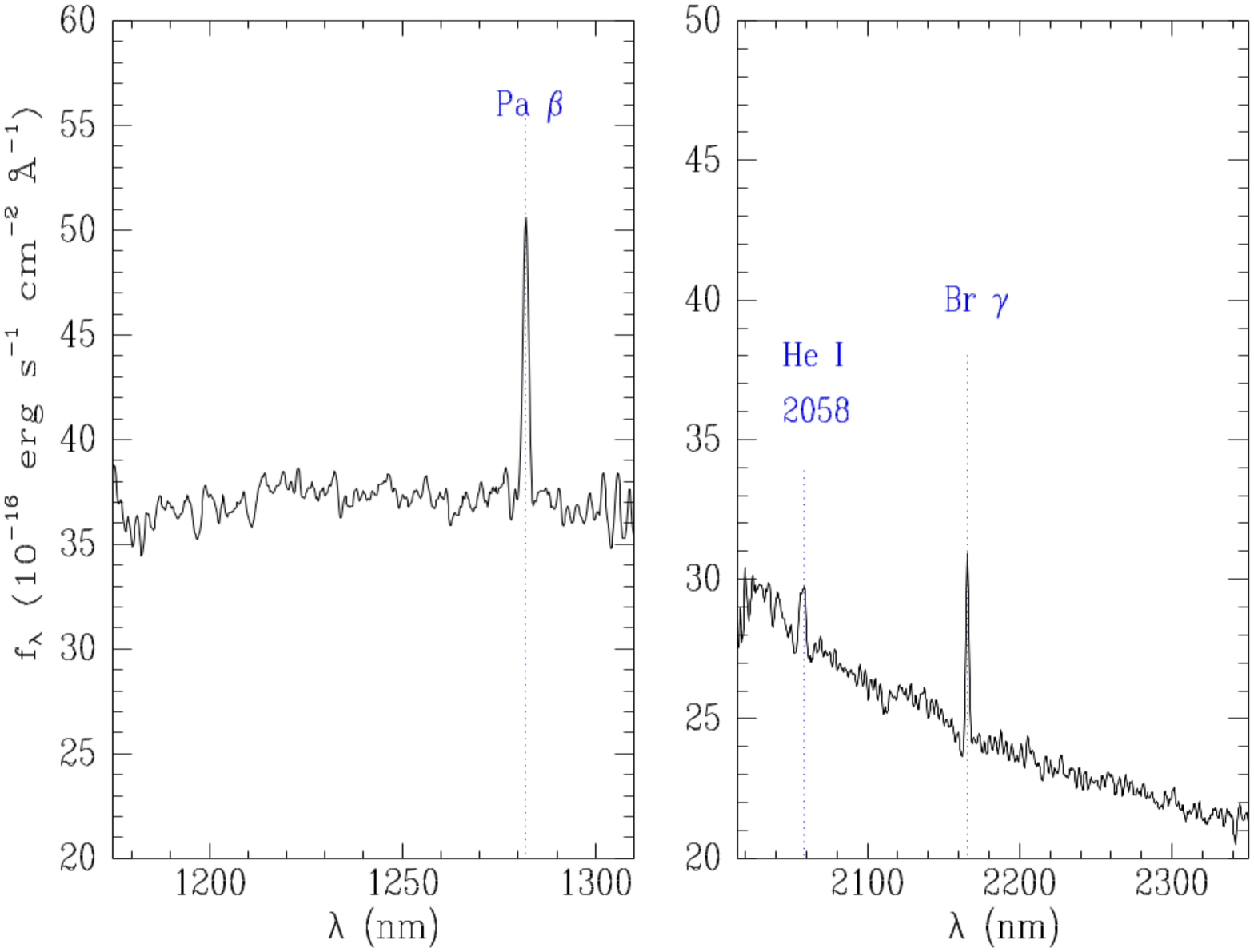}
    \caption{Parts of the J and K band spectra with the prominent emission lines obtained with NOTCam on 2021 Apr 25.}
    \label{fig:notcam}
\end{figure}

\subsection{Optical and Near-Infrared Spectroscopy}\label{subsec:spec}

The optical spectrum of \target\ obtained with {\it RTT-150} on Nov 24, 2020 in frames of the follow-up campaign exhibited a very reddened featureless continuum with only a H$\alpha$ line in the emission. Two additional {\it RTT-150} spectra with the higher resolution of $\sim5.5 \AA$ measured on Nov 25, 2020 with the exposure of 3600 s each allowed us to measure FWHM of the H$_{\alpha}$ line as $\simeq13 \AA$ and equivalent width EW $\simeq - 14 \AA$. Taking into account above mentioned instrumental profile, the intrinsic width of the H$_{\alpha}$ line can be estimated as $\simeq12 \AA$ ( 550  km/s).
This value corresponds to $V_{\rm sin}i \simeq 275$ km/s, that is agreed with the rotational speed of the equatorial disk around Be star.

Later, in the LRIS and Tspec spectra (Figure~\ref{fig:spec}), we identify strong emission lines of \ion{H}{I}, \ion{He}{I}, and \ion{Fe}{II}. We also see strong interstellar absorption lines such as \ion{Na}{I}, \ion{K}{I}, and diffuse interstellar bands (DIBs). These indicate a stellar spectrum of Be type \citep{Gray2009}. There is very little flux below 5000\,\AA\ due to the high extinction, preventing an accurate spectral type from being assigned. However, the presence of strong emission in the \ion{H}{I} Paschen series (Pa7 to Pa19) identifies the star as no later than B2 \citep{Andrillat1988}. In the NIR, we observed the \ion{H}{I} Brackett series (Br7, Br10 to Br24), as well as \ion{He}{I} at $\rm 1.083\,\mu m$, $\rm 1.700\,\mu m$, and $\rm 2.058\,\mu m$. These emission lines signifies a star earlier than B2.5 \citep{Hanson1996}. The non-detection of \ion{He}{I} at $\rm 2.1116\,\mu m$ and singly-ionized helium is indicative of stars later than O9 \citep{Hanson1998, Hanson2005}. The H$\alpha$ profile (inset, upper panel of Figure~\ref{fig:spec}) shows a shallow absorption core on top of an emission line with kinematically broadened full-width half-maximum (FWHM) of $460\,{\rm km\,s^{-1}}$. This value is slightly lower than ones estimated from {\it RTT-150}, but still compatible with typical rotational speeds of the decretion discs around Be stars and signifies a viewing angle that is not close to edge-on \citep{Hanuschik1996}. Taken together, we conclude that \target contains a star with a spectral type of B0-B2e.

We produced a summed spectrum using the LRIS and ESI data, and measured the equivalent width ($EW$) of interstellar absorption lines. Using the relations reported by \cite{Munari1997} and \citet{Yuan2012}, the measured $EW{\rm (}$\ion{K}{I}\,$\lambda 7699)=0.47 \pm 0.01$, $EW{\rm (DIB\,\lambda5780)}=1.73 \pm 0.06$, and $EW{\rm (DIB\,\lambda 6283)}=3.67 \pm 0.04$ can be converted to $E(B-V) = 2.07\pm 0.03$, $2.84\pm 0.10$, and $2.91\pm 0.03$, respectively. We note that the highly reddened nature of \target requires extrapolation of the nominal relations beyond the calibrated ranges. Therefore, the derived $E(B-V)= 2$--3 should be considered as rough estimates.

The NOTCam near-IR spectra taken on 2021 April 25, see Fig.~\ref{fig:notcam},
show strong emission lines in Pa $\beta$, Br $\gamma$ and \ion{He}{I} at $\rm 2.058\,\mu m$, the latter quite broad. The continuum flux is lower than that observed with Tspec
four months earlier, while the emission lines are stronger, the Pa $\beta$ EW is -6.9 $\pm$ 0.5 \AA \ and the Br $\gamma$ EW is -5.3 $\pm$ 0.5 \AA \ in the NOTCam spectra, which is
about the double of that measured four months earlier by Tspec.
From the $K$ band acquisition image we used differential photometry with 9 comparison stars to calibrate roughly the flux and found that the target was 0.89 $\pm$ 0.08 mag brighter than its cataloged 2MASS magnitude.

Now, using the derived constrains on the stellar class of the optical star we can compare observed magnitudes of the IR-counterpart from the 2MASS catalog\footnote{https://cdsarc.unistra.fr/viz-bin/cat/II/246} ($H=11.904\pm0.021$; $Ks=11.481\pm0.018$) with the absolute ones of B0-2e stars. According to \citet{2015AN....336..159W}, the average intrinsic color of such stars is $(H-Ks)_0\simeq-0.04$. Thus, under the assumption of the standard extinction law \citep{1989ApJ...345..245C}, we can estimate the magnitude of the absorption $A_{Ks}$ towards the source as $A_{H}-A_{Ks} = (H-Ks) - (H-Ks)_0$, which gives $A_{Ks}\simeq 0.85$ and $E(B-V)\simeq2.5$.

Correcting the observed magnitude of the counterpart in the $Ks$-filter for the absorption and comparing unabsorbed magnitude with the absolute one, we can estimate \citep[see][for details]{2015AstL...41..394K} the distance to the object in the range of 4--7.5 kpc, which approximately agrees with the values obtained by \citet{2021AJ....161..147B}.

\section{Conclusions}

In this paper we reported a discovery of a new X-ray pulsar \srga/\srge. The source was found by both instruments, \art and \ero, on board \srg during second and third all-sky surveys. The follow-up campaign with X-ray and optical observatories allowed to reveal pulsations with the period of $\simeq742$ s, hard X-ray spectrum with the exponential cutoff and a number of emission lines (H$_{\alpha}$, \ion{He}{I}, Pashen and Braket series) in the optical and infrared spectra of the companion star. Both these factors as well as the source luminosity of $L_X \simeq 4\times10^{35}$\,\lum, estimated from the X-ray spectrum of the source (Table\,\ref{tab:spe}) and {\it Gaia} distance $\simeq8$ kpc, strongly suggest the BeXRB nature of \srga. Moreover, based on the optical and IR spectral measurements we can restrict a spectral type of the companion star as B0-B2e at an estimated distance of 4--7.5 kpc, that approximately agrees with the {\it Gaia} measurements.

Based on the relative stability of the source flux on the time scale of several months, apparently we are dealing with a new member of the subclass of persistent low-luminosity BeXRB systems \citep[see, e.g.][]{2011Ap&SS.332....1R} presumably accreting from the "cold" accretion disk \citep{2017A&A...608A..17T,2017MNRAS.470..126T}. However historical \xmm data point to the possible variability of the X-ray flux caused by an evolution of the optical companion, that is also seen from the long-term monitoring in optics and IR bands. This joint optical and X-ray variability could be naturally explained in terms of decretion disk loss episode, during which the source was quiescent in X-rays and dimmer in optical-NIR. Later, the replenishment of disk around optical star restarted accretion onto the NS, leading to increase of X-ray flux, and, eventually, to \art\, detection.

A relatively hard X-ray band of the \art\ telescope also makes the measured X-ray flux less dependent on the source intrinsic or Galactic absorption and allows the detection of heavily absorbed sources, which could be missed by soft X-ray instruments. Taking into account that the sensitivity of the \srg\ telescopes are exceeding any previous and currently working surveying instruments, the \srg\ observatory allow us to unveil the hidden population of faint persistent objects including the population of slowly rotating X-ray pulsars in BeXRB. The potential of \srg\ to reach this goal is demonstrated by this paper as well as by the discovery of several other new XRPs in the Magellanic Clouds and our Galaxy (see e.g., \citealt{2020ATel13609....1H,2020ATel13610....1M}; Doroshenko et al. 2021).

\begin{acknowledgements}

This work is based on observations with  {\it Mikhail Pavlinsky} \art, and \ero\ X-ray  telescopes aboard \srg\  observatory. The SRG observatory was built by Roskosmos in the interests of the Russian Academy of Sciences represented by its Space Research Institute (IKI) in the framework of the Russian Federal Space Program, with the participation of the Deutsches Zentrum für Luft- und Raumfahrt (DLR). The \art\ team thanks the Russian Space Agency, Russian Academy of Sciences and State Corporation Rosatom for the support of the  \art\ telescope design and development. The SRG/eROSITA X-ray telescope was built by a consortium of German Institutes led by MPE, and supported by DLR.  The SRG spacecraft was designed, built, launched and is operated by the Lavochkin Association and its subcontractors. The science data are downlinked via the Deep Space Network Antennae in Bear Lakes, Ussurijsk, and Baykonur, funded by Roskosmos. The eROSITA data used in this work were processed using the eSASS software system developed by the German eROSITA consortium and proprietary data reduction and analysis software developed by the Russian eROSITA Consortium.
We also would like to thank the \xmm, \nustar, \nicer and \swiftxrt teams for organising prompt observations.  This research has made use of data, software and/or web tools obtained from the High Energy Astrophysics Science Archive Research Center (HEASARC), a service of the Astrophysics Science Division at NASA/GSFC and of the Smithsonian Astrophysical Observatory's High Energy Astrophysics Division. This work made use of data supplied by the UK Swift Science Data Centre at the University of Leicester. Authors (IFB, MRG, RAS) are grateful to TUBITAK, the Space Research Institute, the Kazan Federal University for their partial support in using RTT-150  (the Russian – Turkish 1.5-m telescope  in Antalya).
Partly based on observations made with the Nordic Optical Telescope, owned in collaboration by the University of Turku and Aarhus University, and operated jointly by Aarhus University, the University of Turku and the University of Oslo, representing Denmark, Finland and Norway, the University of Iceland and Stockholm University at the Observatorio del Roque de los Muchachos, La Palma, Spain, of the Instituto de Astrofisica de Canarias. This work was supported by the Russian Science Foundation (grant 19-12-00423).

\end{acknowledgements}

\vspace{-0.3cm}
\bibliography{srgaj2043_bib}
\vspace{-0.3cm}

\end{document}